\documentclass[aps,prl,twocolumn,superscriptaddress]{revtex4}
\usepackage{graphicx}
\begin{document}

\title{Condensation energy and the mechanism of superconductivity}

\author{Sudip Chakravarty}

\affiliation{Department of Physics and Astronomy, University of
California Los Angeles, CA 90095, USA}

\author{Hae-Young Kee}

\affiliation{Department of Physics, University of Toronto, Ontario, Canada M5S 1A7 }

\author{Elihu Abrahams}

\affiliation{Center for Materials Theory, Serin Physics Laboratory,
Rutgers University,  Piscataway, NJ 08854}

\date{\today}  

\begin{abstract}

Condensation energy in a superconductor cannot be precisely defined if mean-field 
theory fails to hold. This implies that in the case of high temperature
superconductors, discussions of quantitative measures of condensation energy must
be scrutinized carefully, because the normal state is  anomalous and the
applicability of a mean-field description can be questioned. A related issue
discussed here is the precise meaning of a superconducting transition driven by
kinetic as opposed to one driven by potential energy; we argue that this is a
semantic question.

\end{abstract}

\maketitle

\paragraph{Introduction:} In an earlier paper (CKA) \cite{Chakravarty1}, we raised
the issue that the notion of superconducting condensation energy \cite{Shoenberg}
is ill-defined if the transition cannot be described by BCS mean-field theory,
where the system turns into a normal Fermi liquid with no pairing correlations
once the superconducting order parameter vanishes. Another purpose of that paper
was to elucidate the interlayer tunneling theory (ILT) \cite{Chakravarty2}. In
particular, we examined the strong version of ILT proposed by Anderson
\cite{Anderson2,Anderson3}, in which  the entire ``condensation energy" was
ascribed to ILT. This proposal turned out to be at variance with the
$c$-axis penetration depth measurements of Moler {\em et al.} \cite{Moler} in
Tl2201 and was thus falsified. Nevertheless, we were interested in understanding
if it is at all possible that ILT plays an important role in enhancing the
transition temperature, $T_c$, by increasing the bare superfluid density, thus
defining ILT in a weaker sense, as an enhancement mechanism over and above an 
in-plane pairing mechanism \cite{Chakravarty3}. 

CKA also noted that nominally optimally doped Tl2201 has a specific heat peak
\cite{Loram} that could be approximately fitted by a two-dimensional (2D) Gaussian
fluctuation contribution to the free energy. This observation
reflects once again the importance of in-plane pairing correlations and was an
important conclusion of CKA. We then asked if there was a
sensible procedure to subtract the 2D fluctuations and use the remainder of the
free energy to understand the effect of ILT in the weaker sense of enhancement of
the bare superfluid stiffness. This was difficult, as the correctness and the
precision of the specific heat measurements \cite{Loram} were unknown and still
are because the measurements are yet to be reproduced  by a second group. In
addition, it was not clear over what range of temperatures the fluctuation
contributions must be fitted. Of course, the very notion of Gaussian fluctuations
in a 2D superconducting transition cannot be meaningful close to the transition.
Despite these difficulties, an approximate subtraction procedure was used by CKA. 
The result was that the enhancement of the bare superfluid stiffness in Tl2201 was
indeed extremely small. Nonetheless, we believe that it is conceptually important
to perform such subtractions, preferably more accurate ones, to estimate the
effects of ILT.  This is also true for multilayer cuprates,
where superficially ILT seems to be important
\cite{Bernhard}, at least in the weaker sense defined earlier. 

\begin{figure}
\centerline{\includegraphics[scale=0.5]{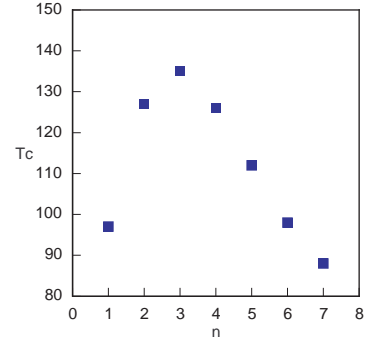}}
\caption{Transition temperature across a homologous series:
HgBa$_2$Ca$_{n-1}$Cu$_n$O$_{2n+2+\delta}$; adapted from Ref.~\cite{Kuzemskaya}} 
\label{fig: Tc}
\end{figure}

To this day, the cause of a striking systematic rise and a subsequent drop in
$T_c$ for a homologous series as a function of the number of layers
in the unit cell is not known. Even if we ascribe the rise to
the enhancement due to ILT
\cite{foot1}, the drop must be ascribed to a competing mechanism that develops
with the increase in the number of layers, perhaps because the inner layers
have a tendency to become underdoped. A homologous series of cuprate
superconductors is a family in which each member has the same charge-reservoir block, 
but $n$ CuO$_2$-planes  in the infinite-layer block, which
consists of $(n-1)$ bare cation planes and $n$-CuO$_2$-planes \cite{Yamauchi}.
Clear systematics of $T_c$ is only evident within a
given homologous series. A well studied example \cite{Kuzemskaya} is the family
HgBa$_2$Ca$_{n-1}$Cu$_n$O$_{2n+2+\delta}$ whose
$T_c$,  optimized with respect  to oxygen concentration, as a function of $n$, is
shown in Fig.~\ref{fig: Tc}.
The formal copper valence $v_{\rm Cu}=2(n+\delta)/n$ is also a bell shaped curve
that peaks at $n=3$. Similar results are known for other families, for which the
transition temperatures follow a similar pattern, often peaking at $n=3$ or $4$. 
The issues of the dependence of $T_c$ on the number of
layers  and the role of ILT remain unresolved. 

We now address the basic question
raised by CKA, namely, is condensation energy a precise quantitative concept for
high temperature superconductors? Given the interest that this subject still
elicits \cite{Bernhard,Aeppli,Demler,Scalapino,Kivelson2,Norman,vDM}, we have
decided to publish this brief note to elaborate further on condensation energy and
on a related theoretical question: can the mechanism of superconductivity be usefully
said to be driven by kinetic as opposed to potential energy? Our conclusions will
be that while it is important to identify the mechanism by which the condensate is
formed, it is a semantic issue as to whether or not we describe the transition as
driven by potential or kinetic energy.

\paragraph{Condensation energy:} Colloquially, the condensation energy is the
difference of the ground state energies between the normal state and the
superconducting state. A little thought reveals several related problems: (1) What
do we mean by the normal state? In particular, what if there are other broken
symmetries \cite{Varma,Fradkin,DiCastro,Chakravarty4,Sachdev} in the regime in which there
is no superconductivity and a further transition to the unbroken symmetry state at
a temperature above the superconducting $T_c$? (2) What if the normal  state
contains superconducting fluctuations? (3) What if the normal state changes as a 
function of the magnetic field, or other tuning parameters used to destroy the
superconducting state? (4) What if the transition to the normal state is not a
first order transition, such that one cannot meaningfully define a notion of a
metastable state that can be accessed in experiments? (5) How should one
correctly  extrapolate the non-zero temperature measurements to $T=0$ to access
the hypothetical normal state with the same set of parameters for which Nature
actually provides us with the superconducting state?
There are indeed simplifying situations, where the complexities mentioned above do
not arise in the practical sense \cite{Shoenberg}. Thus, when mean field theory
holds and the normal state  is a Fermi  liquid with no measurable trace of pairing
correlations, the simplest extrapolation of the normal state below $T_c$ with the
specific heat $C(T)=\gamma T$, where 
$\gamma$ is a constant, is plausible, assuming that there are no other
instabilities of the Fermi liquid at temperatures below $T_c$. One may further
constrain this extrapolation by entropy conservation because the difference of
entropies between the normal state  and the superconducting state is zero at the
mean field $T_c$ and at $T=0$
\cite{Shoenberg}. 

For high temperature superconductors, there are great many complexities.  The
presence of  a pseudogap, quite unlike a BCS superconductor, makes the
extrapolation of the normal state (in fact, even its definition) exceedingly
problematic. If the magnetic field, $H$, is used as a tuning parameter to destroy
superconductivity, its large magnitude, $H>H_{c2}$, may stabilize some other
ordered state \cite{Boebinger}. Moreover, unlike conventional superconductors, for
which the effect of the magnetic field in the normal metal is a weak Landau
diamagnetism, the normal state of high temperature superconductors may not be
so impervious  to such high fields necessary to destroy superconductivity.  The
attempt to destroy superconductivity by doping Zn to replace Cu suffers from
similar problems. In fact, it is empirically known that Zn impurities introduce
magnetic order in high temperature superconductors \cite{Alloul}.

There are even more fundamental reasons for doubting the notion of condensation
energy.  If the transition is a continuous transition, there is no way that one
phase can be continued into the other beyond the transition. Therefore, the
hypothetical normal state cannot exist for the same set of parameters for which
the superconducting state is more stable; the notion of a metastable state is thus
not meaningful for a continuous transition. An exactly solved model illustrates
this point beautifully. Consider the 2D Ising model for which Onsager's result for
the free energy is known for  all temperatures. The analytic continuation of the
free energy, $f_{+}$, from above the ferromagnetic transition point $T_F$ to below
$T_F$ was obtained exactly by Majumdar
\cite{Majumdar}. One gets, close to $T_F$,
\begin{equation}
f_{+}\simeq -\frac{k_BT_c}{4\pi u_F^2}(u-u_{F})^2\left[\ln | u - u_F |+i\pi\right] ,
\end{equation}
where $u=\exp(-4J/k_B T)$, $u_F$ is its value at the 
transition point, $T_F$, and $J>0$ is the ferromagnetic exchange constant. It is
seen that the analytic continuation acquires an imaginary part, which has no
obvious physical meaning. This is true for any continuous transition for which
specific heat exhibits a nonanalytic critical singularity, reflecting a branch
point in the complex plane. It is even true for infinite order transitions, 
as in a six-vertex model. The exact analytic continuation of the free
energy of the six-vertex model was obtained by Glasser {\em et al.}
\cite{Glasser}. If the transition were instead a first order transition, the imaginary
part of the free energy could be interpreted as the decay of the metastable state
\cite{Langer}.

It might be tempting to define condensation energy as the difference between the
exact ground state energy with zero order parameter (unbroken symmetry state) and
the exact ground state energy with a prescribed finite value of the order
parameter (broken symmetry state).  For a broken symmetry with a non-conserved
order parameter, as in a superconductor, this is impossible, simply because the
order parameter and the hamiltonian cannot be simultaneously diagonalized.  To
understand the nature of the broken symmetry state with a non-conserved order
parameter, consider the simplest such case:  an antiferromagnet for which the
staggered order parameter is not conserved. In a bipartite lattice, where the
Marshall sign condition \cite{Marshall} holds, the ground state is always a
singlet. In a finite volume, the symmetry cannot be broken, and, for a large
system, the order parameter will precess slowly so that no orientation is
preferred. The effective hamiltonian, ${\cal H}_{\text {eff}}$, that describes
this precession   depends on the total spin, ${\bf S}_{\text{tot}}$, and  is that
of a rotor, given by
\begin{equation}
{\cal H}_{\text {eff}} = \frac{1}{2\chi}{\mathbf S}_{\text{tot}}^2= \frac{1}{2\chi}S(S+1)
\end{equation}
where $\chi=N\chi_s^{\perp}$ is the total spin susceptibility, in units of 
$g\mu_B,\, \hbar=1$,
$\chi_s^{\perp}$ is the susceptibility per spin with respect to a local uniform
magnetic field oriented perpendicular to the staggered order parameter.  One can
imagine deriving this hamiltonian by a renormalization group analysis,  as the
relevant states are all below the one-magnon state of the smallest non-zero
momentum in a box.  Even though the actual eigenstates are those of total spin,
as $N\to \infty$, a  tower of excited states collapses to the singlet ground
state corresponding to $S=0$, and becomes degenerate with it in the thermodynamic
limit \cite{Anderson}.   The broken symmetry state with a fixed direction of the
staggered order parameter is a coherent superposition in this quasi-degenerate
manifold. Thus, the energetic difference with the singlet ground state vanishes in
the limit $N\to \infty$. The energetic difference between the normal state
and the condensed state is identically zero.

Intuitively, one feels that one should be able to define condensation energy
variationally. Consider two variational wave functions, one of which corresponds
to the superconducting state with broken $U(1)$ gauge symmetry, and the other
corresponding to the normal state. Of course we have to define what we mean by the
normal state--a Fermi liquid, a state with another broken symmetry, etc.
Similarly, we must also define the order parameter symmetry in the superconducting
state. Given a hamiltonian, we can now calculate the expectation value with
respect to these states and find the difference in energy, hence condensation
energy. This is not only model dependent, but also calculation dependent. More
importantly, there is no known experimental method to check the correctness of
this definition of the condensation energy. 

There is one instance in which the condensation energy can be defined with little
ambiguity \cite{Shoenberg}, and that is for a type I superconductor. In this case,
the transition to the normal state as a function of a magnetic field is a first
order  phase transition with only a finite correlation length. If the normal state
is relatively insensitive to the applied magnetic field necessary to destroy
superconductivity, the measurement of the thermodynamic critical field $H_c$, as
$T\to 0$, immediately gives the condensation energy from the formula 
\begin{equation}
G_n - G_s = \Omega\frac{H_c^2}{8\pi},
\end{equation} 
where $G$ is the Gibbs free energy, and $\Omega$ is the volume of the sample.
Unfortunately,  this is unusable for high $T_c$ superconductors because they are
of type II.

\paragraph{Frustrated kinetic energy:} An idea that has been discussed often is
that the superconductivity in the cuprates is driven by the saving of the
electronic kinetic energy in the superconducting 
state \cite{Wheatley,Chakravarty2,Hirsch,Chakravarty1,Chakravarty3,Kivelson}. 
There are some experiments  \cite{Bernhard,vDM2,Bontemps} that could be
interpreted in this manner.

Cuprates are complex materials with intricate  electronic structure. If we
assume that electron-phonon interactions do not play a major role, the problem is
entirely electronic in nature. For concreteness, let us assume that a single band
two-dimensional  Hubbard model is a good effective hamiltonian to understand the
low energy  properties, including the superconductivity of these materials. Even
if the electronic hamiltonian were more complicated, it would make no difference
to our basic argument. For example, we could also incorporate electron-phonon
interaction at the expense of making the discussion more complex. The one-band
Hubbard model describes processes smaller than energy $U$ and is
\begin{equation}
H_{\rm eff}=-t\sum_{\langle ij\rangle}\left(c^{\dagger}_{i\sigma}c_{j\sigma}+\mbox{h.
c.}\right)+U\sum_i n_{i\uparrow}n_{i\downarrow}
\end{equation}
The  higher energy  processes are assumed to be adiabatically  decoupled from the lower
energy processes.  Here $c_{i\sigma}$ is an electron destruction operator of spin
$\sigma$, and $n_{i\sigma}$ is the corresponding density operator.

When $U$ is large, the model can be reduced to the effective hamiltonian called
the $t$-$J$ model, which is
\begin{eqnarray}
H_{t-J}=&-&t\sum_{\langle ij\rangle}\left(c^{\dagger}_{i\sigma}c_{j\sigma}+\mbox{h.
c.}\right)\nonumber \\
&+&J\sum_{\langle ij\rangle}\left({\bf S}_i\cdot{\bf
S}_j-\frac{1}{4}n_in_j\right),
\end{eqnarray}
with $J=4t^2/U$, together with the constraint $n_i\le~1$. The operators $c_{i\sigma}$
still satisfy the fermion anticommutation rule, but one must constrain the Hilbert
space.  This can be done by examining the eigenvalue of a local operator $n_i$.

The $J$ term is a reflection of the frustrated kinetic energy at the level of the
Hubbard model \cite{Anderson4} in the $U\to \infty$ limit, but  at the level of the
$t$-$J$ model, the $J$ term cannot be properly defined to be  kinetic energy: it
does not represent motion of the particles described by the fermion operators.
Moreover, it is neutral under gauge transformation, because both
${\bf S}_i$ and
$n_i$ are. In contrast, the $t$-term is the kinetic energy; it picks up a  Peierls
phase under a gauge transformation and the constraint, being local, remains
unchanged. Thus, it {\em is} meaningful to ask which term plays a more important
role if the superconductivity is described by the $t$-$J$ model, but it is pure
semantics to try to pin the mechanism down as being driven by  kinetic as opposed
to potential energy. What is potential energy at one level is kinetic at the
other. If the $t$-$J$ model is not adequate to describe superconductivity, we must
return to the Hubbard model, and the partitioning of the kinetic and potential
energies will be different.

It is  useful to examine the BCS theory of superconductivity for which the
effective hamiltonian  is the reduced hamiltonian. A textbook calculation shows
that the kinetic energy is increased in the superconducting state, $\delta
(KE)=(\Delta^2/V)[1-N(0)V/2]$, while the potential energy is
lowered,
$\delta (PE)= -\Delta^2/V$, due to the attraction of electrons mediated by
phonons. Here $\Delta$ is the superconducting gap, $V$ the magnitude of the
attractive interaction, and $N(0)$ is the density of states at the Fermi energy.
Although the phonon exchange is  a kinetic process, its  effect is correctly
described  as a potential energy at the level of the reduced hamiltonian. The
increase of the kinetic energy is not in the least surprising because BCS
superconductivity develops on top of a Fermi liquid in which the kinetic energy is
diagonal and unfrustrated. Therefore, it must necessarily be  increased in the
superconducting state. An interesting corollary  is that if superconductivity is
due to the lowering of the electronic kinetic energy in a suitable
low energy effective hamiltonian, it could not develop on top
of a Fermi liquid state, for in a Fermi liquid  the kinetic energy operator is
diagonal; the normal state will have to be a non-Fermi liquid. We may have a new
class of superconductors, but it is still semantics to say  that it is driven by
kinetic energy, for it will surely depend on the low energy effective hamiltonian,
in which a part can appear as a potential energy, which could be a reflection of
frustrated kinetic energy at the preceding level.

We acknowledge
support from the  National Science Foundation, Grant No. 
NSF-DMR-9971138 (S. C.) and Grant No. NSF-DMR-9976665 (E. A. ).   H. -Y.
K.  acknowledges support from the Canadian Institute for
Advanced Research, Canada Research Chair, and Natural Sciences and
Engineering Research Council of Canada.  We also thank T. Geballe for interesting comments.

\end{document}